\documentclass[manuscript]{aastex}
\usepackage{psfig}

\slugcomment{Accepted to the {\it PASP}, July 2002}
\shortauthors{Howell et al.}
\shorttitle{WX Cet}

\begin{document}

\title{Measuring the Boundary Layer and Inner Accretion 
Disk Temperatures for WX Ceti During Superoutburst\footnote{Some observations
were made with the Apache Point Observatory
(APO) 3.5m telescope, which is owned and operated by the
Astrophysical Research Consortium (ARC).}}

\author{Steve B. Howell}
\affil{Astrophysics Group, Planetary Science Institute, Tucson, AZ  85705 \\
howell@psi.edu}

\author{Robert Fried}
\affil{Braeside Observatory, P. O. Box 906, Flagstaff AZ 86002 \\
captain@asu.edu}

\author{Paula Szkody}
\affil{Department of Astronomy, University of Washington, Seattle, WA 98195
\\ szkody@alicar.astro.washington.edu}

\author{Martin M. Sirk}
\affil{Space Sciences Laboratory, UC Berkeley, CA 94720 \\
sirk@ssl.berkeley.edu}

\author{Gary Schmidt}
\affil{Department of Astronomy, University of Arizona, Tucson, AZ 85706 \\
gschmidt@as.arizona.edu}

\begin{abstract}
We obtained EUV photometry, optical spectroscopy, and multi-color 
optical photometry for WX Cet during its 1998 November superoutburst.
WX Cet is only the second short-period, low mass transfer CV 
(TOAD) to ever be observed in the EUV.
Our determined superhump period is consistent with that found
by Kato et al. (0.059 d) and we confirm that superhumps are grey in the
optical.
The optical spectra provide direct evidence that the line emission region
is optically thick and our 
multi-wavelength photometric measurements are used to determine the
inner accretion disk and boundary layer temperatures during superoutburst. 
Using a determined distance to WX Cet of $\sim$130 pc, we 
find T$_{ID}$= 21,000K 
and T$_{BL}$$\sim$72,500K. Both values are in good agreement with that 
expected by models of the superoutburst continuum being produced by the inner
disk and boundary layer.
\end{abstract}


\section{Introduction}

WX Ceti is a member of the Tremendous Outburst Amplitude Dwarf novae (TOAD)
class of cataclysmic variable (CV), with close relatives being systems 
such as WZ
Sge, VY Aqr, and TV Crv. These ``WZ Sge" stars are distinguished by their
properties of 1) large amplitude ($>$ 6 mag) superoutbursts 
(with most TOADs never having normal dwarf novae type
outbursts), 2) periodic photometric signatures during superoutburst termed
superhumps which begin days after maximum 
and establish the orbital period of the binary
to within a few per cent, and 3) mass donors that are likely to be 
sub-stellar (i.e., $\la$0.06 M$_{\odot}$). See Howell \& Skidmore (2000)
for a brief review of the TOADs.

One of the most important properties to measure for a CV is its orbital period.
Photometric observations made during superoutburst offer a means to estimate
P$_{orb}$ for CVs which may otherwise be too faint for, or as yet have no
detailed spectroscopic study. For WX Cet,
Kato et al. (2001) used photometric observations of WX Cet during superoutburst
to determine the superhump
period at 0.$^d$05949. This superhump period is 2.1\%
longer than the orbital period for WX Cet (P$_{orb}$=0.$^d$05827) 
(Rogoziecki \&
Schwarzenberg-Czerny 2001) determined using minimum light observations.
This period difference is typical of CVs which show superhumps
as it measures the difference between
the eccentric superoutbursting accretion disk and the true orbital period
and is related to the CV mass ratio (Warner 1996).

Howell et al. (1999) showed that multi-$\lambda$ superoutburst 
data for four TOADs and three other CVs can be explained by a
``two-temperature" model, from which the temperatures of the inner 
accretion disk and boundary layer can be estimated. 

In this paper, we present our observations of WX Cet during its November
1998 superoutburst. Our measured superhump period agrees with the better 
defined one by Kato et al (2001) and our multi-color optical time-series 
light curves confirm that superhumps are grey in the optical.
Using the determined colors and fitting the optical data, we estimate the
temperature of the inner accretion disk and adding in the EUV photometry, 
we estimate the boundary layer temperature as well.

\section{Observations}

WX Cet went into superoutburst on 1998 November 10 UT. 
During this superoutburst, WX Cet
reached a maximum V magnitude of $\sim$11.8 as estimated by the AAVSO
observers (Figure 1). The observational coverage of the maximum ended near 25
November 1998. On Figure 1 we show the times of our other observations
described below.

\subsection{Braeside Multi-color Optical Photometry}

Optical photometry, in Johnson $UBVRI$ filters, was obtained at Braeside
Observatory (BO) in Flagstaff, Arizona on 1998 November 14-17 UT.
The Braeside telescope is a 0.4-m Cassegrain reflector and used a Loral 422 CCD
as the detector for these observations.
All nights were photometric and the exposure times were 45 s 
($U$ \& $I$), 25 s ($B$), and 20 s ($V$ \&
$R$) for each night with typical 1$\sigma$ differential magnitude
errors of 0.005 mag or less
as determined through use of the statistical tests presented in Howell et al.
(1988).
The primary comparison star was located about 1 arc min SE of
WX Cet. Using the USNO and HST guide star catalogues, we can
estimate the spectral type of this comparison star (F2V) and its optical 
magnitudes: U=10.83, B=10.80, V=10.50, R=10.20, I=10.03.

The BO multi-color observations were taken in a cyclic manner, $VRB$ on the 14th
and $UBVRI$ on the remaining nights. Figure 2(a-d) shows the four nights 
of optical photometry. We have used the comparison star magnitudes 
to place the differential values on an absolute scale with estimated
uncertainties of $\pm$0.1 mag.

The superhumps observed in each color show a variety of shapes and amplitudes
when examined in detail. The shape in all cases is non-sinusoidal and changes 
within a single color in a single night are common. Measurement of the 
trough-to-peak amplitude of the superhumps 
shows a slightly larger amplitude towards the red ($\bar{\Delta U}$=0.11 mag:
$\bar{\Delta I}$=0.17 mag) and a slight overall decrease in amplitude 
from the 14th to the
17th. Some of the superhumps, such as those on the night of the 16th, show
secondary humps similar to, but not as pronounced as those seen in TV Crv
(Howell et al. 1996).

\subsection{{\it EUVE} Photometry}

WX Cet was observed by the {\it EUVE} satellite from 12 to 15 November 1998
UT. WX Cet was a faint target for {\it EUVE} 
so only imaging observations are usable.
The EUV imager (Sirk et al. 1997) 
covers approximately 70-120 \AA~ with a central wavelength 
of $\sim$91\AA. 
The {\it EUVE} observations extended for 3.5 days with a 
total on-source exposure time of
67378 s at a mean count rate of $\sim$0.005 cts/sec. 
Photon event lists (x,y,t) produced from the EUV imager
were used these to perform aperture photometry on the WX Cet data.
Due to the low count rate the data were summed into 0.03 day time bins 
(approximately one-half of the WX Cet orbital period) 
in order to provide reasonable statistics. 
In contrast to the case of T Leo 
(Howell et al. 1999), the EUV light curve of WX Cet,
shown in Figure 3, appears unmodulated. 
Phasing these data on WX Cet's orbital and superhump periods also showed
no modulation.

The EUV count rate observed for WX Cet corresponds to an observed 
flux of 6.23 $\times$
10$^{-16}$ ergs/sec/cm$^2$/\AA~over the wavelength range 
70-120\AA~(see Sirk et al. 1997). 
WX Cet has no published distance estimate, however 
using the Sproats et al. (1996) relation between 
superoutburst amplitude and M$_V$ at minimum and WX Cet's V$_{min}$=17.6, 
we can
determine that M$_V$(min)$\sim$11.5-12.5. This range suggests a
distance to WX Cet of 130$\pm$30 pc. Searching 
the {\it EUVE} data archive for sources near WX Cet we find four main sequence
stars close in location (1.2 to 1.6 degrees)
and distance (105-190 pc) which have estimated
values of their line-of-sight ISM column density 
(N$_H$). The range of the column density values towards WX Cet as determined by 
the ISM indicator stars is 
20.04 $\la$Log N$_H$ $\la$20.41. Calculation of the optical depth at 100
\AA~yields $\tau$=3 to 7 (ISM through-puts of 0.05 to 0.0009),
giving corrected WX Cet EUV fluxes of 1.3 $\times$ 10$^{-14}$ to 
6.9 $\times$ 10$^{-13}$ ergs/sec/cm$^2$/\AA. 

\subsection{Optical Spectroscopy}

A spectrum of WX Cet was obtained on 14 November 1998 17:10 UT, closely
following outburst maximum.  The observations made use of the Steward
Observatory 2.3~m Bok telescope and CCD Spectropolarimeter (Schmidt, Stockman,
\& Smith 1992) in a non-polarimetric configuration.  The data cover the
$4000-8000$\AA\ region at a 3-pixel resolution of 12\AA, and show a steep blue
continuum with broad Balmer and He~I $\lambda4471$ absorption lines, 
all with weak emission cores (Figure 4). Optical
magnitudes were estimated by numerically convolving the spectral data with the
Johnson filter pass-bands, and yield the results: $V=12.3$, 
$B-V=-0.04$, and $V-R=-0.2$.  These are consistent with, but about 0.1~mag
fainter across the board, than our contemporaneous BO multi-color 
photometry, a likely
result of slit losses at the spectrograph.  Half-widths of the absorption lines
measure $2500-3000$ km~sec$^{-1}$, while the emission cores are resolved at
$\sim$900 km~sec$^{-1}$.  Line strengths of the cores show a significantly
flatter Balmer decrement than optically thin recombination, suggesting that
they are formed, at least in part, in an optically thick gas. Since the
continuum and the lines an the accretion disk at outburst are both 
believed to be
optically thick (i.e., leading to weak Balmer absorption), we suggest that
the emission cores are formed in or near the boundary layer in a region that is
thick in the lines but thin in the continuum.

Optical spectroscopy was also obtained by us 
on the 13th and 23rd of November 1998 UT. 
These spectra were obtained
using the Apache Point Observatory (APO) 3.5m telescope and 
Double Imaging Spectrograph with a 1.5 arcsec
slit. The APO spectrum obtained on the
13th is similar to that obtained on the 14th as discussed above.
The spectra from the 23rd are shown in Figure 5 revealing only H$\alpha$
in emission and a rising blue continuum. 
These spectra show that WX Cet
still contained an optically thick accretion disk even at 13 days past maximum
light.

\section{Results}

\subsection{Superhump Period}

Our 4 nights of optical photometry allow an estimate of the superhump period 
of WX Cet.
Three ``methods" were used; (1) performing period
searching (using the phase dispersion minimization (PDM) routine (Stellingwerf
1978))
on each data set separately, each combined color dataset, and
the total color-combined dataset, (2) each dataset
with multiple humps was phased on all the likely periods found in (1), 
and (3) using each dataset
with multiple humps, we simply measured the period between the humps.
The non-sinusoidal, varying amplitude and shape of the superhumps as well as
our limited time coverage did not allow us to determine a highly precise value
for the superhump period. Additionally, we found no period differences
in any of the different color datasets on any of the nights. 
Our 1200 measurements provided a best
fit superhump period of 16.786$\pm$0.1 cycles/day or 0.059 days (85.79 min).
This period is in agreement with the much better determined value of 0.05949(1)
days from Kato et al. (2001).

\subsection{Multi-color Optical Light Curves}

In order to look for any color effects in the amplitude and shape of the
superhumps, we produced light curves for the colors $U-B, B-V, 
V-R, \&  R-I$. 
Figure 6 shows the results with the top plot displaying the simultaneous
$V$ light curve for comparison. The night of the 16th seems to show some
possible color dependence in $U$ and $I$ related to the superhumps, but 
this 
effect is not seen in the data from the 15th (Fig. 6a) or the 17th (not shown).
The curves in Fig. 6 show that within the uncertainties of our measurements,
the superhumps appear to 
be grey over the optical region of the spectrum.

Our observed colors are consistent with those 
observed during the 1989 superoutburst of WX Cet 
by O'Donoghue et al. (1991) 
even though that superoutburst reached a
maximum brightness one magnitude 
brighter than the one under study here.
Assuming an optically thick source, the measured colors for WX Cet correspond
to a color temperature of $U-B$ = 18,000K, $B-V$ \& $V-R$ = 21,000K and 
$R-I$ = 25,000K and we assume this emission represents
the accretion disk during superoutburst.

\subsection{Multi-Wavelength Observations}

Only a few short-period CVs have been observed in the EUV during superoutburst
(Mauche 2002) and only two TOADs; T Leo (Howell et al. 1999) and WX Cet.
Thus, the observation of WX Cet in the EUV provides an unusual opportunity to 
determine the temperature of the inner accretion disk regions.
Howell et al. (1999) showed that using optical (and UV) 
data alone, the temperature of
the cooler inner accretion disk region can be approximated 
by a single blackbody, while 
shorter wavelength observations allow the boundary layer (i.e., the accretion
disk -- white dwarf interface) temperature to be determined. 
In general, the two temperatures differ by $\sim$0.5
dex for short period CVs with the boundary layer temperatures 
being near 80,000K during superoutburst.

Using seven CVs observed during (super)outburst, Howell et al. (1999) 
developed a
model for the temperatures present in the inner accretion disk and 
at the boundary layer (BL)
for CVs, essentially showing that both temperatures decline with orbital
period: The TOAD's inner accretion disk temperature falling more
rapidly than
other short period CVs. The cooler inner accretion disk temperatures for 
the TOADs
are taken to be indicative of an initial (pre-superoutburst) accretion disk
with a optically thin or altogether absent inner accretion disk.
Using the model developed in Howell et al., we calculate the expected inner
disk and boundary layer superoutburst temperatures for WX Cet.

Figure 7 presents the optical ($UBVRI$) and EUV photometric points for WX Cet
for the mid-{\it EUVE} observation time, HJD 2451133. The range for the EUV
photometry is that allowed by the range in estimated ISM column density 
to WX Cet with the ``star" symbol representing the probable value for the EUV flux
(2.51 $\times$ 10$^{-14}$ ergs/sec/cm$^2$/\AA) based on the 
expected ISM column at a distance of 130 pc (log N$_H$ = -13.6).
The inner accretion disk temperature in WX Cet, ``observed" by optical (and UV) 
measurements, is predicted (using Howell et al. 1999) to be 21,000K. 
We already noted that the optical
colors suggest a temperature near 21,000K, in agreement with that expected from
the Howell et al. model. In Fig. 7 we plot a 21,000K blackbody curve (dotted line) which fits
the optical data fairly well. The BO $U$ filter is non-standard and 
its calibration uncertainty may account for the $U$ flux not agreeing with the
model fit. A much cooler accretion disk, produced to fit $U$ as well as the
redder colors, would have to be larger than the entire binary to produce the
needed flux based on our distance estimate.
We also see that the 21,000K blackbody curve falls
short of fitting the EUV photometric point, a result found for all CVs by
Howell et al. (1999). 

To estimate the boundary layer temperature expected for 
WX Cet during superoutburst, we again use the model in Howell et al. and 
calculate a value of T$_{BL}$=72,500K for WX Cet. We plot
the blackbody approximation to the boundary layer in Fig. 7. 
To scale the two blackbody curves in Fig. 7 we proceeded as follows.
Observational determinations of the extent of the boundary layer 
during outburst generally show it to
have an emitting area similar in size to that of the white dwarf (see Howell et
al. 1999 and references therein). Theoretical arguments agree, suggesting that 
the BL may be an equatorial belt
around the WD or through viscous spreading, it may cover a large part of 
the WD surface (e.g., Kutter \& Sparks 1989). 
Given these estimates, we have taken the boundary layer in WX Cet during
superoutburst to have a size which is approximately that of the white dwarf.
We have also assumed
a 0.6 M$_{\odot}$ white dwarf (see Szkody et al. 2002)
with a radius of 8 $\times$ 10$^{8}$ cm.

Using the T$_{BL}$ fit with the assumption that R$_{BL}\sim$R$_{WD}$
and using its best scaled fit to the optical data, our blackbody model
yields a distance to WX Cet
of 120 pc, close to our original estimate.
Scaling the blackbody approximation of the inner accretion disk such that the
sum of the two curves provides a ``by eye" best fit to the optical data,
we find that the ratio of the two (assumed uniform) emitting areas 
has a value of 6.3. Transforming this
scaling into an emitting area, we find an inner disk radius of 
2 $\times$ 10$^{9}$ cm or 2.5 R$_{WD}$. 
Given the range in ISM column, temperatures of 68,000K to 90,000K are strictly
allowed for the boundary layer, however a value of 70,000K is the 
``best fit" to the most likely
value for the EUV flux from WX Cet and temperatures higher than 78,000K
are inconsistent with the optical data and the assumed white dwarf properties.
We note here that it is unlikely that the boundary layer can be adequately
fit by a single blackbody. The assumptions of a single temperature, uniformly
emitting, optically thick boundary layer are likely to be overly simplistic but
given the data in hand, more detailed modelling is not possible. Our
value for T$_{BL}$ should thus be taken to represent an approximation only.

\section{Discussion}

Our multi-color optical and EUV photometric observations were used to 
estimate the superhump period of WX Cet and to show that the humps are color
independent. Our superhump period (0.059 d) agrees with the better
determined value presented in Kato
et al. (2001). Using simultaneous EUV and optical observations, we
fit them with a ``two-temperature" (super)outburst model developed by Howell et
al. (1999). The results indicate that during the peak of the superoutburst, 
WX Cet had a boundary layer temperature near 72,500K and 
an inner accretion disk radius of 2.5 R$_{WD}$ with a temperature near
21,000K.

The above results for WX Cet (and their general agreement with those of T Leo
and other short orbital period dwarf novae),
allow us to provide an estimation of the 
state of the accretion disk during superoutburst in a low mass transfer
dwarf novae (TOAD). The inner disk radius,
being larger than that of the white dwarf, implies that in between the
optically thick inner disk and the white dwarf surface there will be a
rareified region probably at high temperature. While the boundary layer
temperature is $\sim$2-4 times less in the TOADs than in longer 
period dwarf novae (which have higher accretion rates and likely higher mass
primaries; see Howell
et al. 1999), the amount of material accreting onto the white dwarf
is considerably less ($\sim$5 times less; Howell et al. 1995). Therefore, the 
hot boundary layer can form an optically thin region 
($\sim$Str\"ogrem sphere) near the white dwarf which would then be 
a source of coronal emission. 
In fact, a 60-160\AA~{\it Chandra} spectrum taken
during the recent superoutburst of the TOAD WZ Sge (Kuulkers et al. 2002)
showed just this type of coronal spectrum: 
a forest of narrow, bright emission lines with no continuum. 

WX Cet and T Leo are the only TOADs yet observed in the EUV during
superoutburst and we see that the model developed for the inner disk and
boundary layer temperatures appear to work, at least for these 
two systems. The {\it Chandra} observation of WZ Sge was the first X-ray
spectrum of a TOAD obtained during superoutburst. 
Future EUV - X-ray observations of TOADs during superoutburst
are needed in order to provide additional observational tests of the 
proposed model.
 
\acknowledgments
The authors wish to thank the members of the AAVSO who contributed observations
during this superoutburst campaign.
SBH acknowledges partial support for this research from NSF grant 
AST-98-10770 and an AO-7 {\it EUVE} mini-grant and GDS acknowledges NSF grant
AST 97-30792 for support of CV research.

\section{References}

Howell, S.B., Mitchell, K.J., and Warnock, A., 1988, AJ, 95, 247

Howell, S. B., Reyes, A. L., Ashley, R., Harrop-Allin, M. K., and Warner, B.,
MNRAS, 1996, 282, 623

Howell, S. B., et al., 1999, PASP, 111, 342

Howell, S. B., \& Skidmore, W., 2000, NewAR, 44, 33.

Kato, T., Matsumoto, K., Nogami, D., Morikawa, \& Kiyota, S., 2001, PASJ, 53,
893

Kutter, S. \& Sparks, W., 1989, ApJ., 340, 985

Kuulkers, E., Knigge, C., Steeghs, D., Wheatley, P., \& Long, K., 
2002, in {\it The Physics
of Cataclysmic Variables and Related Objects}, B. G\"ansicke, K. Beuermann,
and K. Reinsch Eds., ASP Conf. Series, in press

Mauche, C., 2002, in ``Continuing the Challenge of EUV Astronomy: Current
Analysis and Prospects for the Future, Eds. S. howell, J. Dupuis, D. Golombek,
F. Walter, \& J. Cullison, ASP Conf. Series, in press

O'Donoghue, D., Chen, A., Marang, F., Mittaz, J. P. D., Winkler, H., and
Warner, B., MNRAS, 1991, 250, 363

Rogoziecki, P. \& Schwarzenberg-Czerny, A., 2001, MNRAS, 323, 850

Schmidt, G.D., Stockman, H.S., \& Smith, P.S. 1992, ApJ,
398, L57

Sirk, M. M., Vallerga, J., Finley, D., Jelinsky, P., \& Malina, R., 1997, ApJS,
110, 347

Sproats, L. N., Howell, S. B., \& Mason, K. O., 1996, MNRAS, 282, 1211

Stellingwerf, R. F., 1978, ApJ, 224, 953
 
Szkody, P., G\"ansicke, B., Sion, E.,\& Howell, S. B., 2002,
in {\it The Physics
of Cataclysmic Variables and Related Objects}, B. G\"ansicke, K. Beuermann,
and K. Reinsch Eds., ASP Conf. Series, in press

Warner, B., 1996, "Cataclysmic Variables", Cambridge University Press,

\newpage

Figure captions:

Fig 1: The AAVSO light curve of 
WX Cet during its 1998 November superoutburst. The times of our observations
are marked: A=APO spectra, S=Steward observatory spectrum, EUV = {\it EUVE}
Observation, and Opt = BO multi-color photometry. The insert shows a
close-up view of one of the white light time-series data sets.

Fig 2a: Braeside U band photometry of WX Cet obtained during 1998 Nov. 15-17 UT.

Fig 2b: Braeside B band photometry of WX Cet obtained during 1998 Nov. 14-17 UT.

Fig 2c: Braeside V band photometry of WX Cet obtained during 1998 Nov. 14-17 UT.

Fig 2d: Braeside R band photometry of WX Cet obtained during 1998 Nov. 14-17 UT.

Fig 2e: Braeside I band photometry of WX Cet obtained during 1998 Nov. 15-17 UT.

Fig 3: EUV light curve of WX Cet obtained with the {\it EUVE} satellite 
on 1998 Nov. 12-15 UT.

Fig 4: Low resolution spectrum of WX Cet obtained on 1998 Nov. 14 UT. Note the
steep blue continuum and the Balmer series in absorption blue-ward of H$\alpha$.

Fig 5: High resolution APO spectra of WX Cet obtained on 1998 Nov. 23. Note the
presence of emission cores in H$\beta$ and H$\gamma$ as well as in He I
4471\AA.

Fig 6a: $U-B$, $B-V$, $V-R$, and $R-I$ light curves for 1998 Nov. 15.
The top panel shows the V light curve for reference.

Fig 6b: $U-B$, $B-V$, $V-R$, and $R-I$ light curves for 1998 Nov. 16.
The top panel shows the V light curve for reference.

Fig 7: Multi-wavelength photometry and blackbody fits to the superoutburst 
observations of WX Cet. The vertical range in the EUV photometric value is a
result of the uncertainty in the ISM neutral hydrogen column to WX Cet and the
``star" symbol represents the most likely value.
The two temperature model represent the inner accretion disk (21,000K) 
and boundary layer (72,500K) regions during superoutburst. See text for
details.

\newpage

\begin{figure}
\psfig{figure=fig1.ps}
\end{figure}

\begin{figure}
\psfig{figure=fig2a.ps}
\end{figure}

\begin{figure}
\psfig{figure=fig2b.ps}
\end{figure}

\begin{figure}
\psfig{figure=fig2c.ps}
\end{figure}

\begin{figure}
\psfig{figure=fig2d.ps}
\end{figure}

\begin{figure}
\psfig{figure=fig2e.ps}
\end{figure}

\begin{figure}
\psfig{figure=fig3.ps}
\end{figure}

\begin{figure}
\psfig{figure=fig4.ps}
\end{figure}

\begin{figure}
\psfig{figure=fig5.ps}
\end{figure}

\begin{figure}
\psfig{figure=fig6a.ps}
\end{figure}

\begin{figure}
\psfig{figure=fig6b.ps}
\end{figure}

\begin{figure}
\psfig{figure=fig7.ps}
\end{figure}

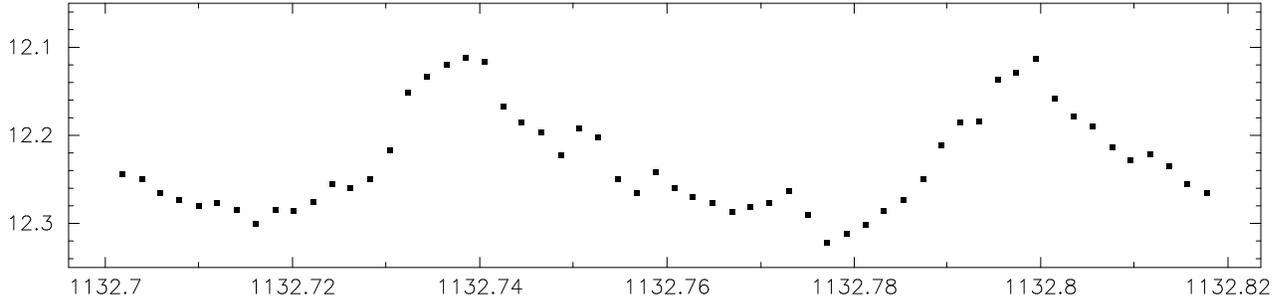
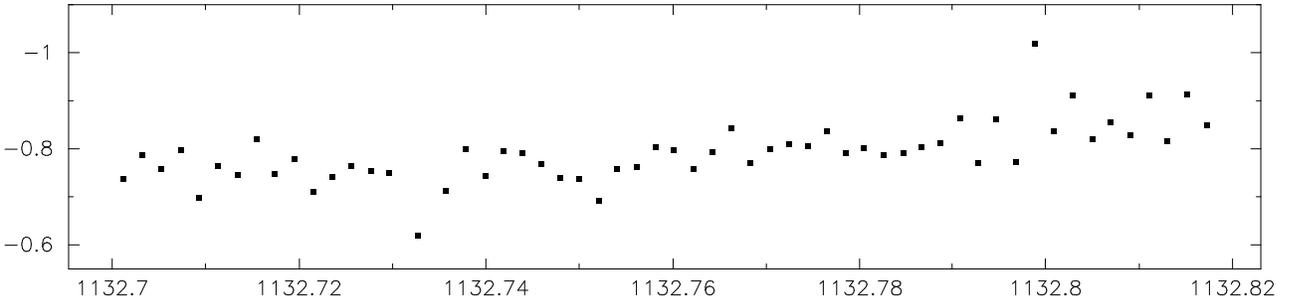
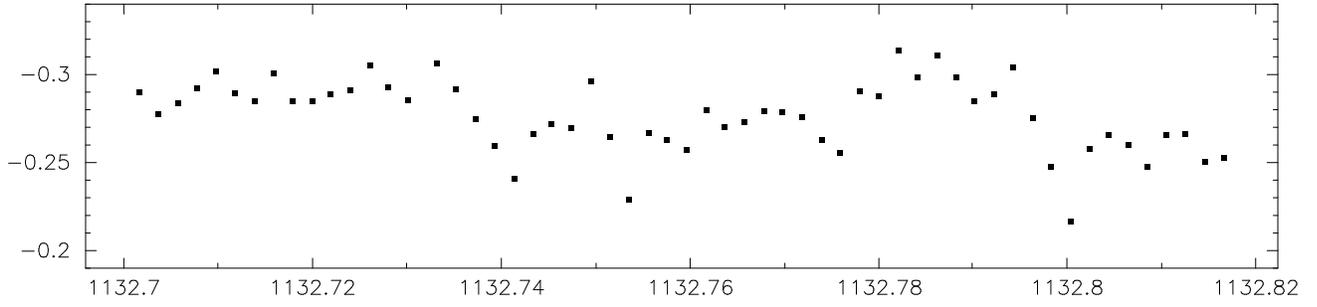
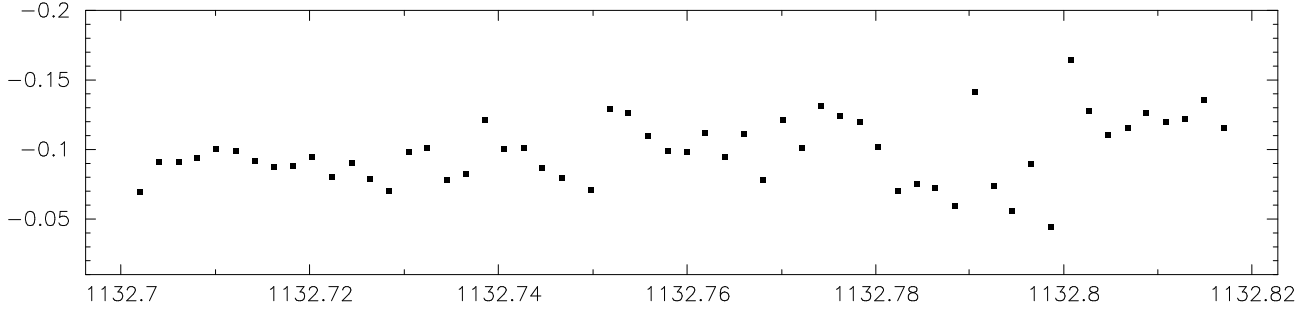
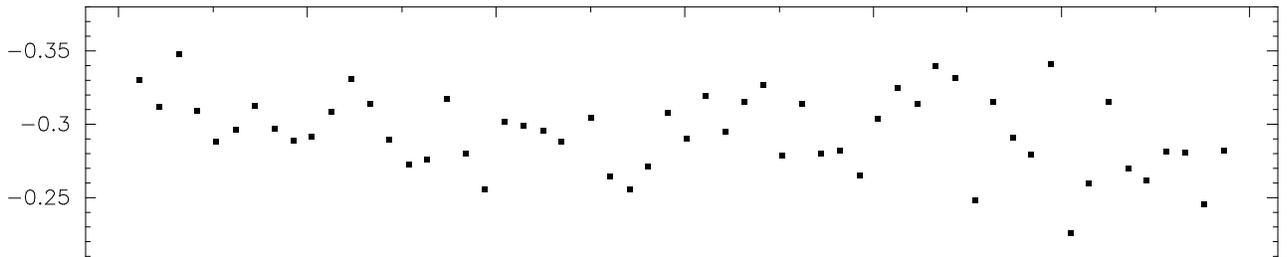
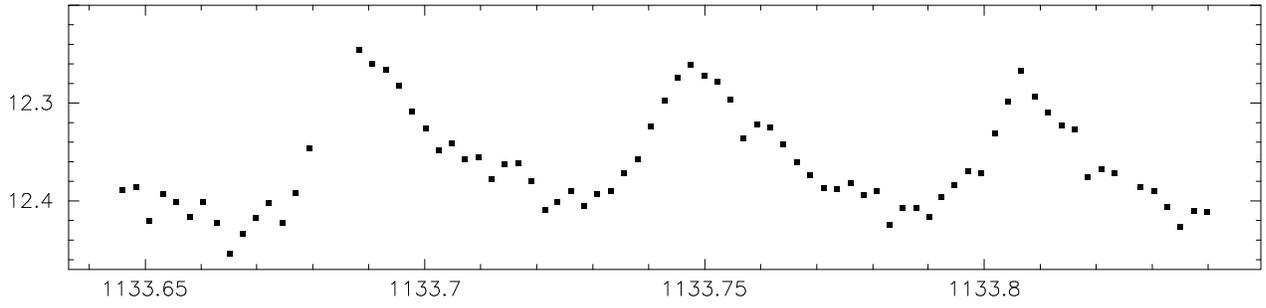
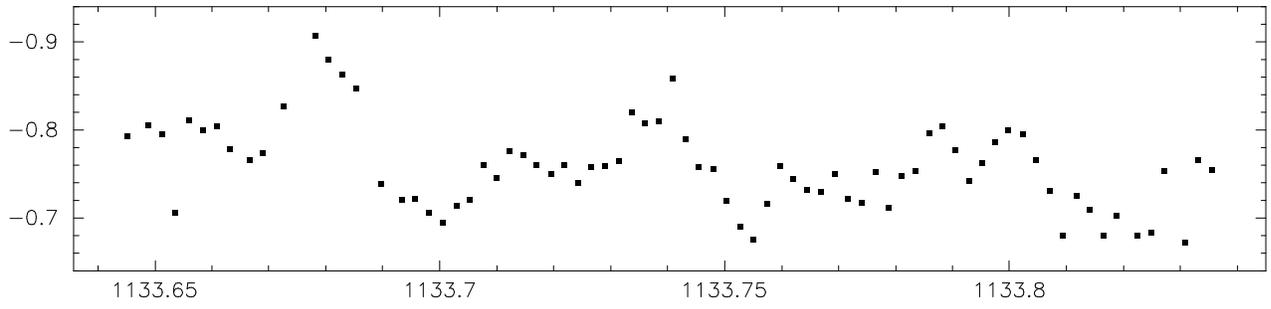
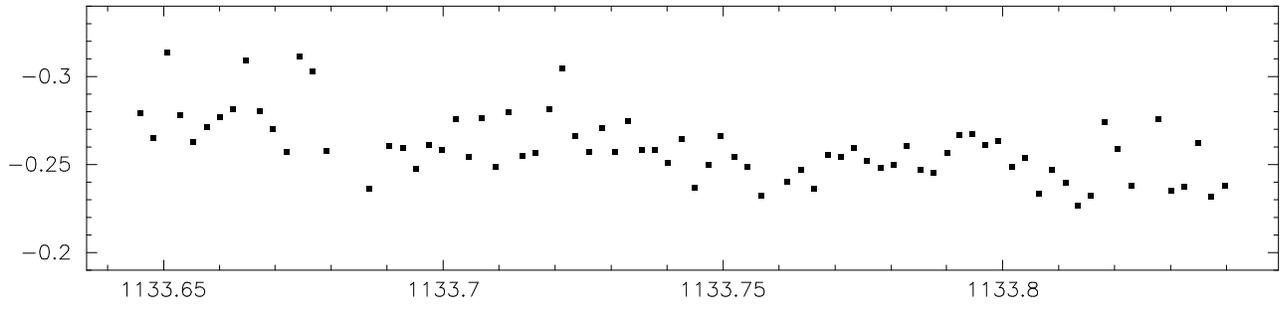
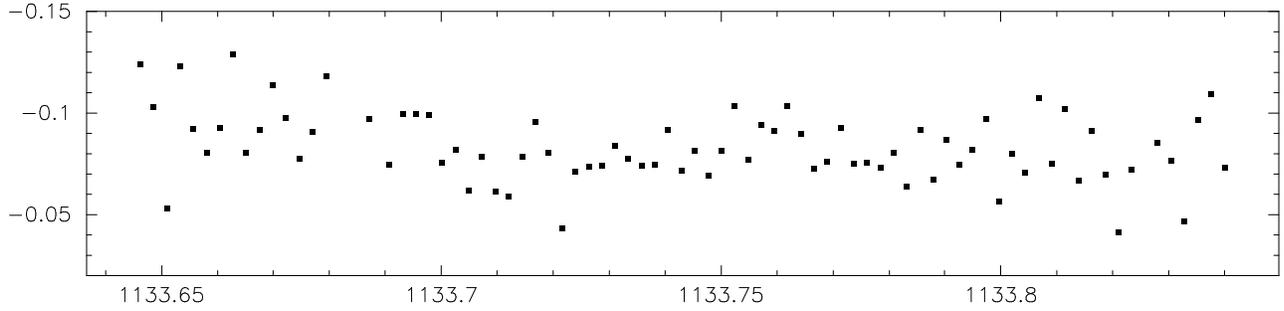
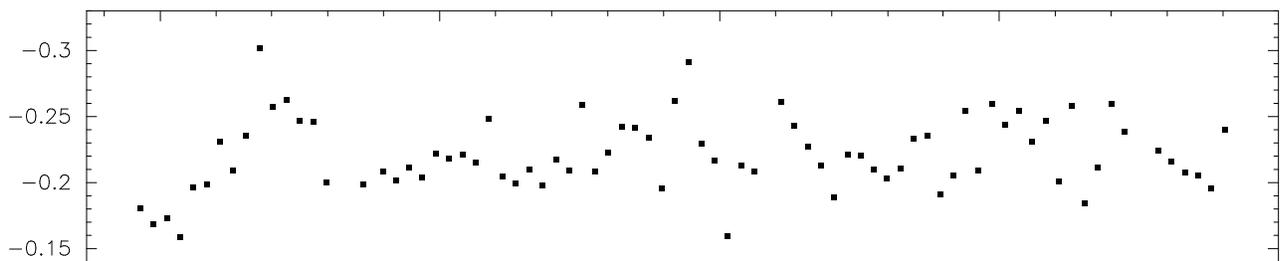
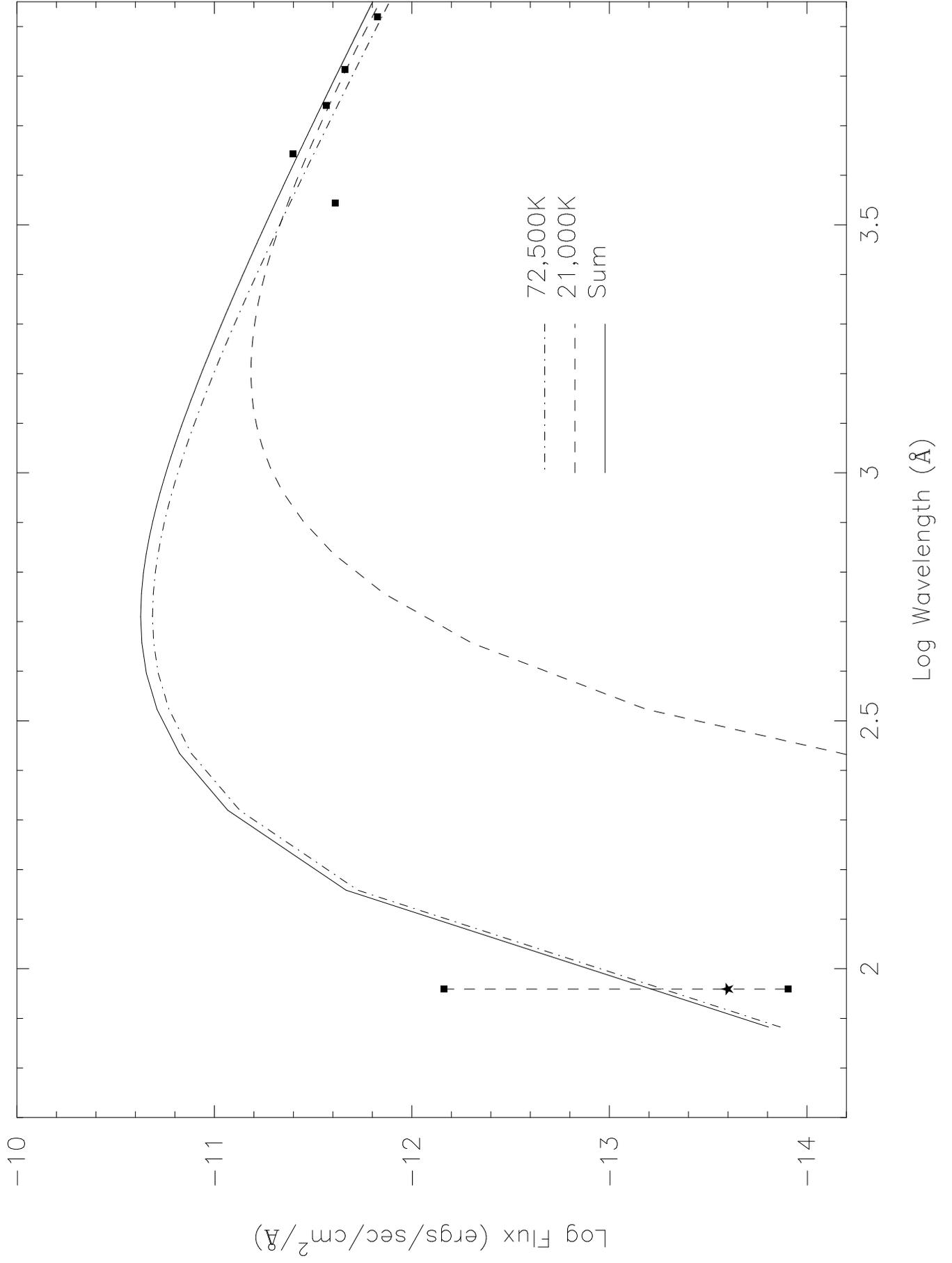
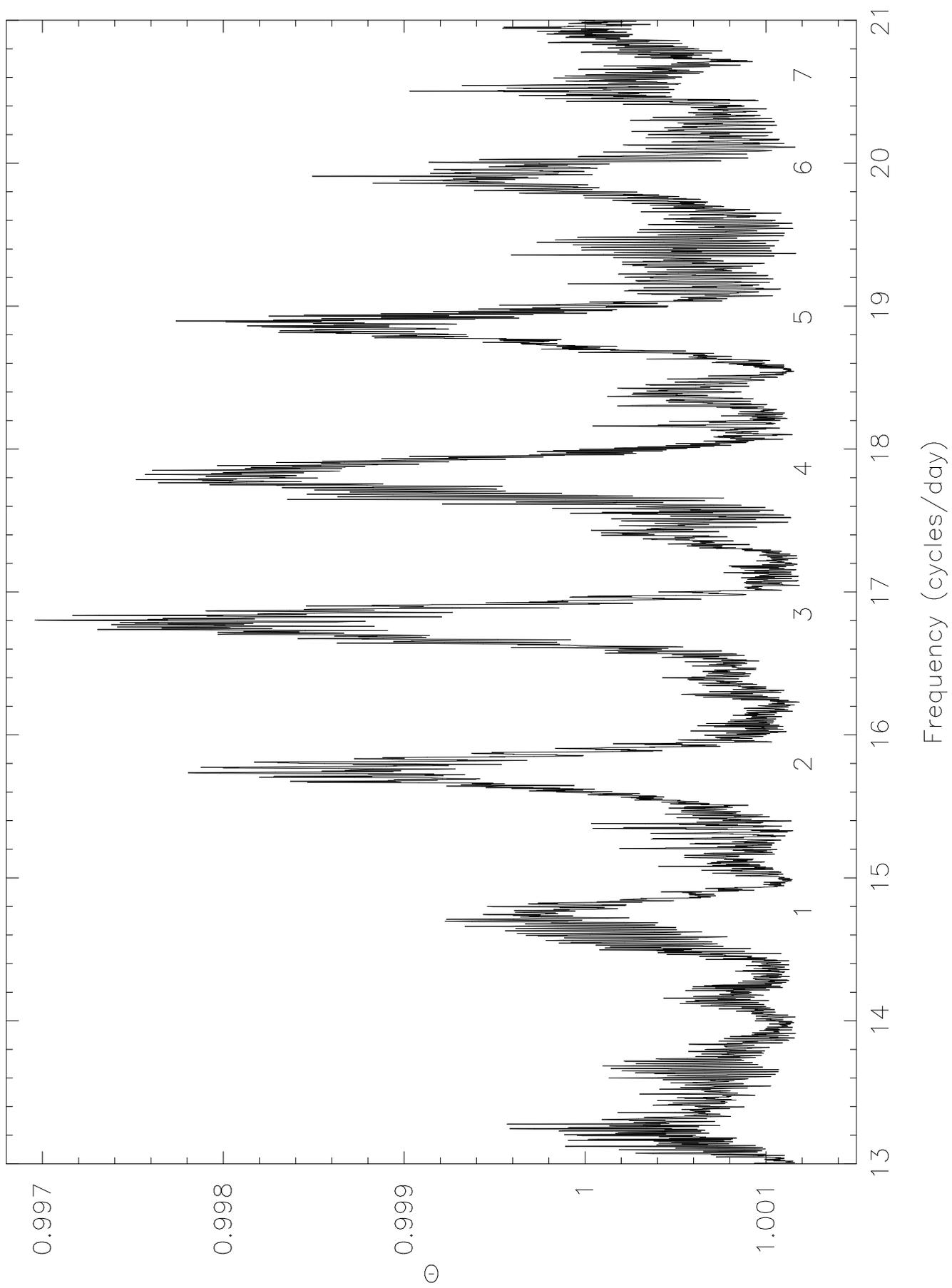
\end{document}